\documentclass[prl,showpacs,twocolumn,preprintnumbers]{revtex4}

\usepackage{graphicx}
\usepackage{dcolumn}
\usepackage{amsmath}
\usepackage{bm}

\begin{document}

\title{Overcoming the Rayleigh Criterion Limit with Optical Vortices}

\author{F. Tamburini}

\author{G. Anzolin}

\author{G. Umbriaco}

\author{A. Bianchini}

\author{C. Barbieri}

\affiliation{Department of Astronomy, University of Padova, vicolo
dell' Osservatorio 2, Padova, Italy. }

\begin{abstract}

We experimentally and numerically tested the separability of two
independent equally--luminous monochromatic and white light sources
at the diffraction limit, using Optical Vortices (OV), related to
the Orbital Angular Momentum (OAM) of light. The diffraction pattern
of one of the two sources crosses a phase modifying device
(fork--hologram) on its center generating the Laguerre--Gaussian
(L--G) transform of an Airy disk. The second source, crossing the
fork--hologram in positions different from the optical center,
acquires different OAM values and generates non--symmetric L--G
patterns. We formulated a criterion, based on the asymmetric
intensity distribution of the superposed L--G patterns so created,
to resolve the two sources at angular distances much below the
Rayleigh criterion. Analogous experiments carried out in white light
allow angular resolutions which are still one order of magnitude
below the Rayleigh criterion. The use OVs might offer new
applications for stellar separation in future space experiments.

\end{abstract}


\maketitle

\textit{\textbf{Introduction.} ---} Electromagnetic (EM) radiation
carries momentum: the linear momentum is usually associated with
radiation pressure, while the angular momentum is associated with
the polarization of the optical beam~\cite{al92}. However, the
transfer of angular momentum of light to material
bodies~\cite{beth36,he95} together with the creation of OVs have
been demonstrated in several basic experiments, also in white
light~\cite{lea03}. OAM of light cannot be ascribed only to a
multipole component of the wavefront along the propagation
direction~\cite{pad95,al00}, but is also a quantum property of
single photons~\cite{arl99,MT02,oneil02}. OAM has also attracted
attention for its applications in quantum
communication~\cite{mai01,vaz02,arl98} and in
astronomy~\cite{har03,tam06,maw05,foo05}. The use of OV has recently
allowed the separation of two monochromatic sources with 95\%
intensity contrast~\cite{lee06}.

The generation of beams carrying OAM proceeds thanks to the
insertion in the optical path of a Phase Modifying Device (PMD)
which imprints a certain vorticity on the phase distribution of the
incident beam~\cite{arl98}. Consider a monochromatic pencil of
light, described by an EM wave propagating along the $z$ direction
of a reference frame $S$ and with transversal amplitude distribution
$u(\mathbf{r})$. The PMD is placed on the $(x,y)$ plane and its
optical singularity is centered on the origin $O$. When the incident
beam has its propagation axis centered on the active region of the
PMD, its Airy diffraction pattern is transformed into a series of
donought--shaped concentric rings which do not carry net transverse
momentum relative to their symmetry axis. Light beams carrying OAM
are described by L--G modes for which the EM field amplitude is
\small
\begin{multline} \label{eqn:lgmode}
 u_{lp}^{L-G}(r, \varphi, z) = \sqrt{\frac{2 p!}{\pi (p + l)!}}
 \frac{1}{w(z)} {\left[\frac{r \sqrt{2}}{w(z)}\right]}^l \\
 \times \, L_p^l \left[\frac{2 r^2}{w^2(z)}\right] \exp
 \left[\frac{-r^2}{w(z)^2}\right] \exp \left[\frac{-\mathrm{i} k
 r^{2}}{2R(z)}\right] \\
 \times \, \exp \left[-\mathrm{i} (2 p + l + 1) \arctan
 \left(\frac{z}{z_R}\right)\right] e^{-\mathrm{i} l \varphi }
\end{multline}
\normalsize
where $z_R$ is the Rayleigh range of the beam, $w(z)$ the beam waist
radius (the spot size), $L_p^l (x)$ an associated Laguerre
polynomial and $R(z)$ the radius of curvature. The azimuthal and
radial indices $l$ and $p$ provide information about the OAM of the
beam and the number of radial nodes of the associated intensity
profile, respectively. As a consequence of the phase factor $\exp
(\mathrm{i} l \varphi )$, any cylindrically symmetric L--G mode
carries an OAM per photon, relative to its symmetry axis, equal to
$l \hbar$. The intensity distribution of an L--G mode with $p = 0$
is, in polar coordinates, $I(r, \phi) = \frac{2}{\pi w^2
l!}\left(\frac{r \sqrt{2}}{w} \right)^{2 l} \exp \left(\frac{-2
r^2}{w^2}\right)$, while the regions having the same intensity
describe a ring structure. The center of the OV is dark, with a
radius that, depending on the OAM value, reaches a maximum at
$r_{\max} = \frac{\sqrt{2}}{2} w \sqrt{l}$, while the intensity
scales as $I(r_{\max}) = \frac{2}{\pi w^2 l!} l^l e^{-l}$, which,
for large values of the azimuthal index decreases as the inverse of
the square root of $l$~\cite{pad95}. Strongly focused, tilted and
off--axis beams are not described by a single donought--shaped L--G
mode, but show a more complicated asymmetric pattern made by a
superposition of an infinite number of coaxial L--G modes. The OAM
index of such a weighted superposition usually possesses a
non--integer value~\cite{lenz96,hel04,cg03,vas05,vaz02}.

The diffraction images of two point--like sources (Airy disks) are
resolved when the maximum of intensity of one source overlaps the
first intensity minimum of the second equally bright source. The
historical definition~\cite{daco32} refers to a symmetrical
double--peaked profile with a central dip 5\% lower than the
intensity maxima. For a telescope having diameter $D$, at wavelength
$\lambda$, the separation is achieved at  $\theta_R = 1.22 \,
\lambda / D$~\cite{hecht02}. Techniques to overcome this limit have
been developed in certain special cases~\cite{Mu05}. In~\cite{swa01}
the separation of integer--valued L--G modes is obtained by
measuring the distance of the maxima of an Airy disk centered on a
L--G donought mode at distances different than the Rayleigh
criterion. In particular, only for $l = 1$ is achieved a
sub--Rayleigh separation corresponding to 0.64 times
$\theta_R$~\cite{palaphd}. Larger integer values of the azimuthal
index $l$ produced separation angles larger than $\theta_R$.

In this paper, we show that the Rayleigh limit may be better
overcome by using for the two sources a combination of integer and
non--integer values of $l$.

\bigskip

\begin{figure}[!htb]
\centering
\includegraphics[width=6cm, keepaspectratio]{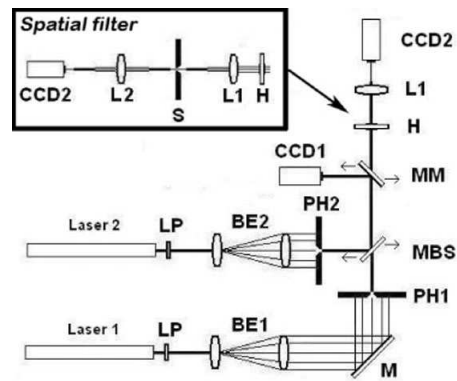}
\caption{Optical setup: LP's are neutral polarizing filters, BE1 and
BE2 beam expanders, M a fixed mirror, MBS a moving beam splitter, MM
a moving plane mirror, L1 a biconvex lens, H the fork--hologram,
CCD1 and CCD2 two CCD cameras. The inset represents the spatial
filtering for the white light: L2 is a camera lens and S a narrow
slit (aperture $\alt$ 1 mm). The two sources were obtained by using
two distinct He-Ne lasers, in the monochromatic case, or a halogen
lamp and two optical fibers in white light (not represented here).}
\label{fig1}
\end{figure}

\begin{figure}[!htb]
\centering
\includegraphics[width=6cm, keepaspectratio]{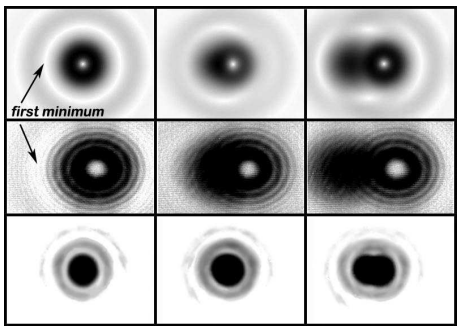}
\caption{Images of the separation of two nearby monochromatic
sources having same intensity (inverted colors). \textbf{Upper row:}
numerical simulations of L--G modes generated by an $l=1$
fork--hologram at different separations. \textbf{Central row:} the
corresponding experimental results. \textbf{Bottom row:} Airy
patterns of the two sources on the hologram plane. \textbf{Left
column:} the superposed sources. \textbf{Mid column:} the sources
separated by $0.42$ times the Rayleigh criterion. \textbf{Right
column:} the sources separated by $0.84$ times the Rayleigh
criterion radius.}
\label{fig2}
\end{figure}

\textit{\textbf{Sub--Rayleigh separability.} ---} In this section we
present the experimental and numerical results, achieved using OVs,
about the sub--Rayleigh separation of two equally intense and
uncorrelated monochromatic sources and also experimental results
obtained in white light. The optical scheme of the experiment is
shown in Fig.~\ref{fig1}. Two $632.8$ nm He--Ne lasers generate the
two independent monochromatic sources, while white light sources
were provided by a halogen lamp and two optical fibers. Airy disks
were produced by equal pairs of pinholes with diameters of $35$,
$50$, $400$ and $500$ $\mu$m placed at the same distance from the
fork--hologram H. During the experiment one beam was always
coinciding with the optical center of the hologram, thus generating
an OV with integer topological charge $l = 1$. The other beam
spanned the hologram in different positions starting from the
optical center. For non--central positions the second beam formed an
OV carrying non--integer components of OAM. This phenomenon is
typical of laterally displaced and angularly deflected beams, which
Airy disk transforms are composed of an infinite set of azimuthal
harmonics expressed either in the form of Bonnet--Gaussian beams or
with well--defined orbital and azimuthal values of L--G
modes~\cite{vas05,zam06}. However, the central dark regions are
always superposed, because they have been generated by the same
central optical singularity of H. The separation of the diffraction
figures of the two sources was obtained by using a moving beam
splitter (MBS). The two beams were kept parallel with a tolerance of
$10^{-5}$ degrees (0.17 $\mu$rad), with negligible effects on the
OAM value due to beam tilting. In white light we corrected the
chromatic dispersion due to the hologram by spatially filtering the
first generated diffraction order with the slit S placed on the
Fourier plane of the achromatic lens L1 (see inset in
Fig.~\ref{fig1}). The hologram H, $20$ lines/mm, has an active area
of $2.6 \times 2.6 \text{ mm}^2$ with a $50\mu$m--sized optical
singularity. It is blazed at the first diffraction order and its
efficiency is about $80\%$ at the laser's wavelength. An incoming
Gaussian beam is projected by H in a superposition of L--G modes
where the dominant modes have $l=0$ and $l=1$ for every value of
$p$~\cite{vaz02}. H was placed perpendicular to the optical axis at
a distance $d = 430$ mm away from the two pinholes, giving a Fresnel
number $F \simeq 0.15$, sufficient to satisfy the Fraunhofer
diffraction prescriptions to obtain Airy diffraction patterns on the
hologram~\cite{hecht02}. This was verified by inserting in the
optical path a moving mirror (MM) at 45 degrees and analyzing the
spots with the CCD1 camera (see Fig.~\ref{fig2}, bottom row). By
measuring the ratio of the distances of the first two diffraction
pattern maxima with respect to the center, we obtained a value of
$1.59$, close to $1.64$ as provided by the theory.

In \cite{swa01,palaphd} was discussed the separability of two
identical overlapping OVs with the same OAM value: the
integer--valued L--G transforms of the two Airy profiles were
numerically computed for the same integer $l$ value and, following
the Rayleigh separability criterion, the OVs of the two beams were
rigidly superposed so that the maximum of one coincided with the
dark center of the other. In this case, a sub--Rayleigh separation
is achieved only when $l = 1$ at an angular distance $\theta_{l = 1}
= 0.64 \; \theta_R$. For OAM values $l > 1$ the size of the OV
becomes larger, scaling with $\sqrt{l}$, and two identical donought
modes are separated at angular distances larger than $\theta_R$. For
example, for $l = 2$, the angular separation would be $\theta_{l =
2} = 1.03 \; \theta_R$, already worse than the Rayleigh criterion
(see lower inset of Fig.~\ref{fig5}).

\begin{figure}[!htb]
\centering
\includegraphics[width=7cm, keepaspectratio]{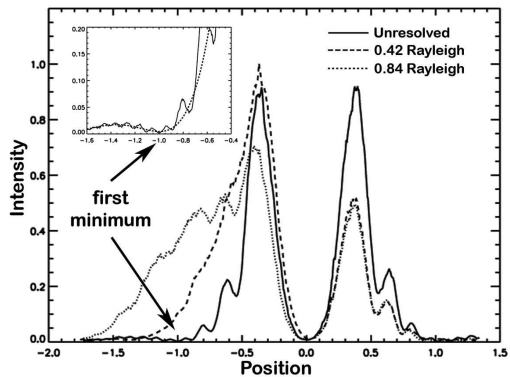}
\caption{\textbf{Main figure:} Experimental intensity profiles of
the superposed L--G modes of the two monochromatic sources
normalized with respect to the peaks relative to coincident sources.
The three cases shown refer to the same separations as in
Fig.~\ref{fig2}. When one of the two sources is shifted to an
off--axis position the combined profile becomes clearly asymmetric.
\textbf{Inset:} Zoom of the position of the \textit{first minimum}
of the L--G transform of the Airy disk; the dotted curve represents
the results of numerical simulations.}
\label{fig3}
\end{figure}

\begin{figure}[!htb]
\centering
\includegraphics[width=7cm, keepaspectratio]{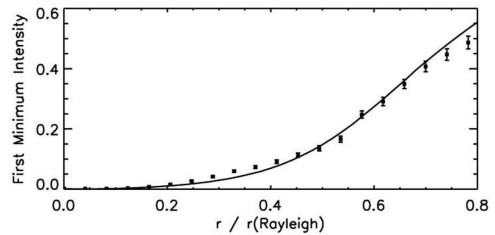}
\caption{The intensity of the \textit{first external minimum} (see
Fig.~\ref{fig3}) of the combined profile produced by two equally
intense sources vs the off--axis shift of one of them in units of
the Rayleigh radius $\delta_R$. The continuous curve is derived from
numerical simulations of the L--G patterns produced by the two
sources for different separations. Dots and error bars refer to our
laboratory results. The intensity is normalized with respect to that
of the two superposed sources.}
\label{fig4}
\end{figure}

\begin{figure}[!htb]
\centering
\includegraphics[width=7.5cm, keepaspectratio]{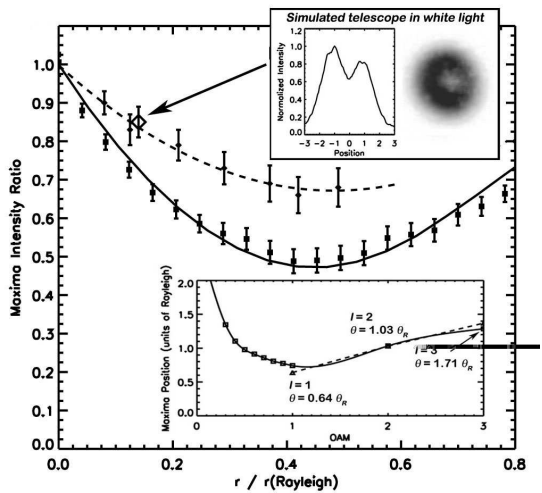}
\caption{\textbf{Main figure:} Ratio between the intensities of the
peaks of the superposed L--G modes vs the off--axis shift of the
spot in units of the Rayleigh radius. The solid line represents the
theoretical expectations for monochromatic light, while the
superposed dots and errorbars are the experimental data. The dashed
line is a linear interpolation of two experimental data obtained in
white light. A 5\% difference (see text) between the intensities of
the peaks implies in both cases a separability at least one order of
magnitude better than the Rayleigh limit. \textbf{Lower inset:}
Positions of the maxima of the L--G modes relative to the OV (in
units of the Rayleigh radius) vs OAM. Triangles show the angular
separation values between two equally charged OVs as calculated
in~\cite{swa01}. \textbf{Upper inset:} White light L--G modes
generated by two equally luminous simulated stars as seen with a
diffraction limited telescope. The angular separation is $\sim 10$
times below the Rayleigh radius (empty diamond in the main figure).}
\label{fig5}
\end{figure}

The experimental results shown in Fig.~\ref{fig2} depict the
sub--Rayleigh separability of two monochromatic OVs produced with
our setup. Using the 400 $\mu$m pinholes, the Rayleigh criterion
limit was $\theta_R = 1.93$ mrad, corresponding to a linear
separation $\delta_R = 834$ $\mu$m on the hologram plane. The upper
row of Fig.~\ref{fig2} shows the numerical simulations of L--G modes
generated by an $l = 1$ fork--hologram, the central row shows our
experimental results and the bottom row shows the corresponding Airy
figures of the two equally intense sources. The first column of
Fig.~\ref{fig2} represents two coincident sources, the second column
represents two sources separated by $0.42 \, \delta_R$ and the third
column shows the sources separated by $0.84 \, \delta_R$.

In Fig.~\ref{fig3} we plot the intensity of the central section of
the combined profiles of the L--G patterns along the direction
connecting the two sources. The profiles are normalized with respect
to the peaks of the superposed sources. We can observe that, as the
linear separation $\delta$ increases, the initially symmetric
profile is distorted and the intensity of the two maxima assume
different values. When $\delta = 0$, the profile is symmetric and
the two maxima have the same intensity given by the exact
superposition of two identical L--G modes. When $\delta = 0.42 \;
\delta_R$, the intensity of one of the two maxima decreases much
faster than the other one, the latter tending to reach the intensity
level of the single source. In the limit $\delta \rightarrow \infty$
the on--axis beam still has $l = 1$, but the off--axis one will
assume $l \simeq 0$. For this reasons we devised a separability
criterion based on the peaks' intensity ratio.

Fig.~\ref{fig4} plots the intensity taken at the \textit{first
external minimum} (as indicated in Fig.~\ref{fig3}) of the combined
profile produced by two equally intense sources vs the separation of
the off--axis source. As the separation increases, the intensity
taken at the position of the \textit{first external minimum}
monotonically increases due to the contribution of the
non--symmetric profile of the off--axis source with non--integer
L--G modes.

In Fig.~\ref{fig5} (main figure) we plot the intensity ratios of the
main peaks produced by separated sources. We tested the separation
of the off--axis monochromatic source in a range $0 \leq \delta \leq
700$ $\mu$m, with a step of 35 $\mu$m. The experimental data show a
good agreement with the theoretical curve obtained from numerical
simulations of the L--G transform of an Airy disk, following the
mode decomposition shown in~\cite{vaz02}. We see that the intensity
ratios reach a minimum value 0.48 when the separation is $\sim 0.42
\; \delta_R$. By roughly assuming the Airy disk as a Gaussian
one~\cite{oemra}, we estimate the non--integer OAM value of the
off--axis source at $\delta = 0.42 \; \delta_R$ to be $l \simeq
0.4$. The lower inset of Fig.~\ref{fig5} reports the positions of
the main peak of the simulated L--G modes in units of the separation
$\delta_R$ vs the estimated OAM values. We see that, considering
only the positions of the intensity maxima, sub--Rayleigh
separations are achieved only for $0.45 < l < 2$, with a minimum
around $0.7 \; \delta_R$, which is still worse than the limit $0.64
\; \delta_R$ found in~\cite{swa01,palaphd}. In the same plot we also
show the separations obtainable with the method discussed
in~\cite{swa01} for two identically charged donoughts with $l = 1,
2, 3$. Instead, if we analyze the relative intensities of the
asymmetric peaks produced by the off--axis object, we may achieve a
more efficient sub--Rayleigh separability limit. Obviously, the
actual separability depends on the S/N ratio of the data. The main
figure also reports the data obtained in white light that suggest a
different slope at small separations with respect to the
monochromatic behavior. We might mimic the historical definition of
the Rayleigh criterion by assuming that two identical sources are
just resolved when the intensities of the asymmetric peaks differ by
at least 5\%. In the monochromatic case, we would reach a
theoretical separability 50 times better than the Rayleigh limit.
Analogously, the results obtained in white light suggest a
separability about 10 times better then the Rayleigh limit. This
lower resolution is mainly due to the non--perfect spatial filtering
and lower degree of coherence of the sources. In the upper inset of
Fig.~\ref{fig5} we show a successful application of the separability
criterion in white light, where we simulated the OVs of a double
star with angular separation $\sim 10$ times below the Rayleigh
limit, as seen with a diffraction limited telescope having the same
focal ratio of the 122 cm Galileo telescope in Asiago.

\bigskip

\textit{\textbf{Conclusions.} ---} In this paper we have shown that
when two sources cross an $l = 1$ fork--hologram and one beam is
always centered with the optical singularity, we obtain different
L--G patterns if the second beam is off--centered because it
acquires a non--integer OAM value producing an asymmetric pattern.
Using this property, we have shown both numerically and
experimentally that both monochromatic and white light OVs can be
used to reach separabilities at least one order of magnitude lower
than the Rayleigh limit. Our results might have interesting
applications in several techniques of applied optics and also in
astronomy, especially in space missions and in telescopes with
adaptive optics.

We would like to thank the Institut f\"{u}r Experimentalphysik,
University of Wien, Zeilinger--Gruppe for support and helpful
discussions and the three anonymous referees for their helpful
comments. This work has been partly supported by ESO and by the
University of Padova.

\end{document}